%% file: HH_PRL_corr.tex
\documentclass[prl,reprint,preprintnumbers]{revtex4-1}

\usepackage{dcolumn}
\usepackage{graphicx}
\usepackage{amsmath,amssymb}
\usepackage{xspace}

\graphicspath{{figures/}}

\input{definitions}

\begin{document}

\preprint{MPP-2016-80, NSF-KITP-16-040, ZH-TH-14/16}

\title{Higgs boson pair production in gluon fusion at NLO\\ with full top-quark mass dependence}

\author{S.~Borowka}
\affiliation{Institute for Physics, Universit{\"a}t Z{\"u}rich, Winterthurerstr.190, 8057 Z\"urich, Switzerland}
\affiliation{Kavli Institute for Theoretical Physics, University of California, Santa Barbara, CA 93106, USA}
\author{N.~Greiner}
\affiliation{Institute for Physics, Universit{\"a}t Z{\"u}rich, Winterthurerstr.190, 8057 Z\"urich, Switzerland}
\author{G.~Heinrich}
\author{S.~P.~Jones}
\author{M.~Kerner}
\author{J.~Schlenk}
\author{U.~Schubert}
\author{T.~Zirke}
\affiliation{Max Planck Institute for Physics, F\"ohringer Ring 6, 80805 M\"unchen, Germany}

\date{\today}

\begin{abstract}
We present the calculation of the cross section and invariant mass distribution for Higgs boson pair production in gluon
 fusion at next-to-leading order (NLO) in QCD. Top-quark masses are fully taken into account
 throughout the calculation. The virtual two-loop amplitude has been
 generated using an extension of the program {\sc GoSam} supplemented with an interface
 to {\sc Reduze} for the integral reduction. The occurring integrals have been calculated numerically using the program
{\sc SecDec}. Our  results, including the full top-quark mass
dependence for the first time, allow us to assess the validity of various
 approximations proposed in the literature, which we also recalculate.
 We find substantial deviations between the NLO result and the
 different approximations, which emphasizes the importance of including the
 full top-quark mass dependence at NLO.
\end{abstract}
 
\pacs{}

\maketitle 

\section{Introduction}

The couplings of the Higgs boson  to electroweak bosons and heavy fermions are being 
established as Standard-Model-like at an impressive rate.  
In contrast, the measurement of the Higgs boson self-coupling, 
which is vital in order to confirm the mechanism of electroweak symmetry breaking, 
is still outstanding, and will have to wait until the LHC high-luminosity upgrade.
However, the Higgs boson self-coupling(s) could be enhanced by physics Beyond the Standard Model (BSM), 
and it is an important task
to be able to distinguish BSM effects from effects due to higher order corrections in perturbation theory.

Gluon fusion is the dominant production channel for Higgs boson pair production. 
However, as this process 
proceeds via a heavy quark loop already at the leading order (LO), the next-to-leading order
corrections involve two-loop four-point diagrams with two masses, $m_h$ and $m_t$, 
and the analytic calculation of two-loop four-point integrals with different internal and external 
 mass scales has not been achieved so far.

The leading order (one-loop) calculation of Higgs boson pair production in gluon fusion 
has been performed in Refs.~\cite{Glover:1987nx,Plehn:1996wb}. 
NLO corrections in the $m_t\to\infty$ limit for both the Standard Model and the MSSM have been performed in
Ref.~\cite{Dawson:1998py}. Finite top-quark mass corrections to the NLO result 
have been calculated in Refs.~\cite{Grigo:2013rya,Frederix:2014hta,Grigo:2014jma,Maltoni:2014eza,Grigo:2015dia,Degrassi:2016vss}.
The NNLO QCD corrections in the $m_t\to\infty$ effective field theory  also have been 
computed~\cite{deFlorian:2013uza,deFlorian:2013jea,Grigo:2014jma}, 
and they have been supplemented by an expansion in $1/m_t^2$ in Ref.~\cite{Grigo:2015dia}.
In the effective field theory, resummation at NLO+NNLL has been considered in Ref.~\cite{Shao:2013bz}, and
recently, even matched NNLO+NNLL resummed results became available~\cite{deFlorian:2015moa}. 
The dominant uncertainty therefore is given by the unknown top-quark
mass effects at NLO.

The top-quark mass effects have been included in various approximations in the literature:
\begin{enumerate}
\item[(i)] The ``Born-improved HEFT (Higgs Effective Field Theory)'' approximation, 
which is the one employed in the program {\sc Hpair}~\cite{Plehn:1996wb,Dawson:1998py}.
It uses 
the heavy top-quark limit throughout the NLO calculation, in combination with a re-weighting factor 
$B/B_{HEFT}$, where $B$ denotes the leading order result in the full theory.
In {\sc Hpair} the re-weighting is done at  matrix element level, 
but after the angular integration of the phase space, 
while in Ref.~\cite{Maltoni:2014eza}
it is done on an event-by-event basis.

\item[(ii)] The ``FT$_{approx}$'' result of Refs.~\cite{Frederix:2014hta,Maltoni:2014eza}
contains the full top-quark mass dependence in the real radiation, 
while the virtual part is rescaled by the re-weighting factor mentioned above.\\
It was found that (ii) leads to a total cross section which is about 10\% smaller than 
the one obtained using Born-improved HEFT.

\item[(iii)] The ``FT$_{approx}^{\prime}$'' result\,\cite{Maltoni:2014eza} is as in (ii) 
for the real radiation part, while 
it uses partial NLO results for the virtual part, specifically, the exact results for the two-loop triangle diagrams 
as far as they are known from single Higgs boson
production~\cite{Graudenz:1992pv,Spira:1995rr,Harlander:2005rq,Harlander:2012pb}.

\item[(iv)] HEFT results at NLO and NNLO have been improved by an expansion in
$1/m_t^{2\rho}$ in
Refs.~\cite{Grigo:2013rya,Grigo:2014jma,Grigo:2015dia,Degrassi:2016vss},
where Ref.~\cite{Grigo:2015dia} contains corrections up to 
$\rho^{\rm{max}}=6$ at NLO, and $\rho^{\rm{max}}=2$ for the
soft-virtual part at NNLO. 
In Ref.~\cite{Grigo:2015dia} it is also demonstrated that the sign of the
finite top-quark mass corrections depends on whether the re-weighting factor
is applied at differential level, \ie before the integration over the
partonic center of mass energy, or at total cross section level. 
\end{enumerate}
All these results suggest that 
the uncertainty on the cross section due to 
top-quark mass effects is $\pm$10\% at NLO.

\vspace*{3mm}

In this letter we present results for the total cross section and the Higgs boson pair 
invariant mass distribution 
for the process $gg\to hh$ at NLO, 
including the full top-quark mass dependence. The analytically unknown two-loop integrals have been calculated numerically 
with the program {\sc SecDec}~\cite{Carter:2010hi,Borowka:2012yc,Borowka:2015mxa}.
Our results settle the long-standing question about the 
uncertainty related to the various approximations which have been calculated so far.

\section{NLO calculation}
\label{sec:calculation}

\subsection{Amplitude structure}
\label{sec:amp}

At any loop order, the amplitude for the process $g(p_1)+g(p_2)\to h(p_3)+h(p_4)$
can be decomposed into form factors as
\begin{align}
&{\cal M}_{ab}=\delta_{ab}\eps_1^\mu\eps_2^\nu\,{\cal
  M}_{\mu\nu}\label{eq:FFdeco}\\
&{\cal M}^{\mu\nu}=F_1(\hat{s},\hat{t},m_h^2,m_t^2,D)\;T_1 ^{\mu\nu}+F_2(\hat{s},\hat{t},m_h^2,m_t^2,D)\;T_2 ^{\mu\nu}\;,\nn
\end{align}
where $\eps_1^\mu,\eps_2^\nu$ are the gluon polarization vectors,
$a,b$ are colour indices, and
\begin{align}
  \hat{s}=(p_1+p_2)^2,\quad \hat{t}=(p_1-p_3)^2,\quad \hat{u}=(p_2-p_3)^2\;.
\end{align}
The decomposition into tensors carrying the Lorentz structure is not unique. With the following definitions
\begin{eqnarray}
T_1 ^{\mu\nu}&=& g^{\mu\nu}-\frac{p_1^{\nu}\,p_2^{\mu}}{p_1\cdot  p_2} \;,\label{eq:Ttensors} \\
T_2 ^{\mu\nu}&=& g^{\mu\nu}+\frac{1}{p_T^2\,(p_1\cdot p_2)}\,\tilde{T}_2 ^{\mu\nu}\;,\nn\\
\tilde{T}_2 ^{\mu\nu}&=&
\left\{
  m_h^2 \,p_1^{\nu}\,p_2^{\mu} - 2\,(p_1\cdot p_3) \,p_3^{\nu}\,p_2^{\mu}- 2\,(p_2\cdot p_3) \,p_3^{\mu}\,p_1^{\nu}\right.\nn\\
  &&\left.+ 2\,(p_1\cdot p_2) \,p_3^{\nu}\,p_3^{\mu}\right\}\;,\nn\\
\mbox{ where } && p_T^2=(\hat{t}\hat{u} - m_h^4)/\hat{s}\;,\nn\\
&&T_1\cdot T_2=D-4\;, \;T_1\cdot T_1= T_2\cdot T_2=D-2\;,\nn
\end{eqnarray}   
we have~\cite{Glover:1987nx}
\begin{align}
{\cal M}^{++}&={\cal M}^{--}=-F_1\;,\;
{\cal M}^{+-}={\cal M}^{-+}=-F_2\;. 
\end{align}
At leading order, we can further split $F_1$ into a triangle diagram
and a box diagram contribution, $F_1=F_\triangle+F_\Box$. 
As the form factor $F_\triangle$ only contains the triangle diagrams, which  
have no angular momentum dependence, it can be attributed entirely to
an s-wave contribution.
The form factor $F_2$ contains only box contributions. At NLO in QCD, the
feature persists that only $F_1$ contains diagrams involving the
triple Higgs coupling.
The form factors $F_1$ and $F_2$ can be attributed to the spin-0 and spin-2 states 
of the scattering amplitude, respectively.

\vspace*{3mm}

We construct projectors $P_j^{\mu\nu}$ such that 
 \begin{eqnarray*}
P_1^{\mu\nu} {\cal M}_{\mu\nu}&=&F_1(\hat{s}, \hat{t}, m_h^2,m_t^2,D)\;, \\
P_2^{\mu\nu} {\cal M}_{\mu\nu}&=&F_2(\hat{s}, \hat{t}, m_h^2,m_t^2,D)\;.
\end{eqnarray*}

For the projectors in $D$ dimensions we can use as a basis the tensors $T_i^{\mu\nu}$
defined in Eqs.~(\ref{eq:Ttensors}).
The projectors  can be written as
\begin{eqnarray}
P_1^{\mu\nu} &=&\quad\frac{1}{4}\,\frac{D-2}{D-3} \,T_1^{\mu\nu}
-\frac{1}{4}\,\frac{D-4}{D-3} \,T_2^{\mu\nu}\;,\label{eq:proj1}\\
P_2^{\mu\nu}&=& -\frac{1}{4}\,\frac{D-4}{D-3} \,T_1^{\mu\nu}
+\frac{1}{4}\,\frac{D-2}{D-3} \,T_2^{\mu\nu}\;.\label{eq:proj2}
\end{eqnarray}

\subsection*{LO cross section}
\label{sec:lo}

The partonic leading order cross section  can be written as
\begin{equation}
\hat \sigma^{\mathrm{LO}}
 =\frac{1}{2^9\,\pi\,\hat{s}^2} \int_{\hat t_-}^{\hat t_+} d\hat t \,
 \left\{ \left|F_1 \right|^2+\left| F_2\right|^2 \right\},
\label{eq:sigmahatLO}
\end{equation}
where
\begin{eqnarray}
 \hat{t}^\pm&=&m_h^2-\frac{\shat}{2}\,(1\mp\beta_h)\;,\;
\beta_h^2=1-4\frac{m_h^2}{\shat}\;.
\end{eqnarray}
The leading order form factors $F_i$ with full mass dependence can be found \eg in Refs.~\cite{Glover:1987nx,Plehn:1996wb}. 

For the total cross section, we also have to integrate over the parton distribution functions, so we have
\begin{equation}
\sigma^{\mathrm{LO}}  =  \int_{\tau_0}^1 d\tau~\frac{d{\cal L}_{gg}}{d\tau}~
\hat\sigma^{\mathrm{LO}}(\hat{s} = \tau s)\, .
\label{eq:sigmalo}
\end{equation}
The luminosity function is defined as
\begin{equation}
\frac{d{\cal L}_{ij}}{d\tau}=\sum_{ij} \int_{\tau}^1{dx\over x} f_i(x,\mu_F) f_j\biggl({\tau\over x},\mu_F\biggr)\, ,
\end{equation}
where $s$ is the square of the hadronic centre of mass energy, $\tau_0 = 4m_h^2/s$,
$\mu_F$ is the factorization scale and $f_i$ are the parton distribution functions (PDFs) for parton type~$i$.

\subsection*{NLO cross section}

The NLO cross section is composed of various parts, which we will discuss separately in the following:
\begin{equation}
\sigma^{\mathrm{NLO}}(pp \rightarrow hh) = 
\sigma^{\mathrm{LO}} + 
\sigma^{\mathrm{virt}} + \sum_{i,j\in \{g,q,\bar{q}\}}\sigma_{ij}^{\mathrm{real}}
\label{eq:sigmanlo}
\end{equation}


\subsection{The virtual two-loop amplitude}
\label{sec:virt}

For the virtual two-loop amplitude, we use the projectors defined in 
Eqs.~(\ref{eq:proj1}),(\ref{eq:proj2})
to express the amplitude in terms of the scalar form factors $F_1$ and $F_2$.

The virtual amplitude has been generated with an 
extension of the program {\sc GoSam}~\cite{Cullen:2011ac,Cullen:2014yla}, 
where the diagrams are generated using {\sc Qgraf}~\cite{Nogueira:1991ex} and then further processed 
using {\sc Form}~\cite{Vermaseren:2000nd,Kuipers:2012rf}. 
This leads to about 10000 integrals before any symmetries are taken into account. 
The two-loop extension of {\sc GoSam}  contains 
an interface to {\sc Reduze}~\cite{vonManteuffel:2012np},
which we used for the reduction to master integrals. We have defined 8 integral families with 9 propagators each.
For the 6 and 7 propagator non-planar topologies we could not achieve
a complete reduction with our available computing resources using the reduction programs {\sc Reduze}~\cite{vonManteuffel:2012np}, 
{\sc Fire}~\cite{Smirnov:2014hma} or {\sc LiteRed}~\cite{Lee:2013mka}. 
In this case we evaluated the tensor integrals directly, exploiting the fact that {\sc SecDec}
can calculate integrals with (contracted) loop momenta in the numerator.

After the partial reduction, we end up with 145 planar master
integrals plus 70 non-planar integrals and a further 112 integrals
that differ by a crossing. 
As the master integrals contain up to four independent mass scales, 
$\hat{s}$, $\hat{t}$, $m_t^2$, $m_h^2$, only a small subset is known analytically.
Therefore we have calculated all the integrals numerically using the
program {\sc SecDec}-3.0~\cite{Borowka:2015mxa}.
We partially used a finite basis~\cite{vonManteuffel:2014qoa} for the
planar master integrals, as far as it turned out to be beneficial for
the numerical integration.

The interface to {\sc SecDec} has been constructed such that the coefficients of the master integrals 
as they occur in the amplitude are taken into account when evaluating the integrals numerically. 
For each integral, once a relative accuracy of 0.2 is reached, the number of sampling points is then set dynamically according to two criteria: 
(i) the contribution of the integral including its coefficient to the error estimate of the amplitude 
and (ii) the time per sampling point spent on the integral.
The numerical integration is continued until the 
desired precision for the full amplitude is reached. 
This procedure allows for a precise evaluation of the amplitude, without spending an unnecessary amount of time 
on individual integrals which are suppressed in the full amplitude.

For the numerical integration we use a quasi-Monte Carlo method based on a rank-one lattice rule~\cite{Li:2015foa,QMCActaNumerica,nuyens2006fast}. 
For suitable integrands, this rule provides a convergence rate of $\mathcal{O}(1/n)$ as opposed to Monte Carlo or adaptive 
Monte Carlo techniques, such as {\sc Vegas}~\cite{Lepage:1980dq}, which converge $\mathcal{O}(1/\sqrt{n})$, where $n$ is the number 
of sampling points. The integration rule is implemented in {\sc OpenCL\,1.1} and a further ({\sc OpenMP} threaded) {\sc C++} 
implementation is used as a partial cross-check. The 
665 phase-space points used for the current publication
were computed with $\sim$16 dual {\sc Nvidia Tesla K20X} GPGPU nodes using a total of
4680 GPGPU hours.

We use conventional dimensional regularization (CDR) with $D=4-2\epsilon$. 
The top-quark mass is renormalized in the on-shell scheme and the QCD coupling in the $\msbar$ scheme with $N_f = 5$. 
The top-quark mass counterterm 
is obtained by insertion of the mass counterterm into the heavy quark propagators.
Alternatively, the mass counterterm can be calculated by taking the derivative of the one-loop amplitude with respect to $m_t$.
We have used both methods as a cross-check.

\subsection{Real radiation}
\label{sec:real}
The contributions from the real radiation, $\sigma_{ij}^{\mathrm{real}}$,  can be divided 
into four channels, according to the partonic subprocesses  $gg\to hh+g,gq\to
hh+q,g\bar{q}\to hh+\bar{q},q\bar{q}\to hh+g$.
The $q\bar{q}$  channel is infrared finite.

We have generated the one-loop amplitudes for all subprocesses
with the program {\sc GoSam}~\cite{Cullen:2011ac,Cullen:2014yla}. 
For the subtraction of the infrared poles, we use the Catani-Seymour dipole formalism~\cite{Catani:1996vz}.
Further we use a phase-space restriction parameter $\alpha$ to limit the subtractions 
to a smaller region in phase space, as suggested in Ref.~\cite{Nagy:2003tz}.
We have retained the full top-quark mass dependence 
throughout the calculation of the $2\to 3$ matrix elements and IR subtraction terms.
For the phase-space integration we use the {\sc Vegas} algorithm~\cite{Lepage:1980dq} as implemented in the {\sc Cuba} library~\cite{Hahn:2004fe}.

The infrared poles of the virtual contribution $\mathrm{d}\hat{\sigma}^{\mathrm{virt}}$ cancel in the combination 
$(\mathrm{d}\hat{\sigma}^{\mathrm{virt}} + \mathrm{d}\hat{\sigma}^{\mathrm{LO}} \otimes \mathbf{I})$,
where the $\mathbf{I}$-operator is given by
\bea
{\bom I}
&=&\frac{\als}{2\pi}\,\frac{(4\pi)^\eps}{\Gamma(1-\eps)}\left(\frac{\mu^2}{\shat}\right)^\eps
\left\{\frac{2 C_A}{\eps^2}+\frac{\beta_0}{\eps} + \mathrm{finite} \right\}\,.\label{eq:Iop}
\eea

\subsection{Checks}

We have checked that for all calculated phase space points the numerical cancellations of the
poles in  $\epsilon$ are within the numerical uncertainties.
For a  randomly chosen sample of phase-space points we calculated the poles with higher accuracy and 
obtained a median cancellation of five digits.

Our implementation of the virtual two-loop amplitude is checked to be invariant under the interchange of $\hat{t}$ and $\hat{u}$ 
by recomputing 10 randomly selected phase-space points. The part of the amplitude known from single Higgs boson production 
is checked against the program of Ref.~\cite{Harlander:2012pb}. Further, the one-loop amplitude is computed 
using an identical framework to the two-loop amplitude and is checked against the result of Ref.~\cite{Glover:1987nx}.

We have verified the independence of the amplitude from the phase space restriction parameter $\alpha$.
Further, we have compared to the results of Ref.~\cite{Maltoni:2014eza} for the approximations (i) and 
(ii) mentioned above, 
and found agreement within the numerical uncertainties~\cite{YR4}.

As a further cross-check we 
have also calculated mass corrections as an expansion in $1/m_t^{2}$ in the following way:
we write the partonic differential cross section as
\begin{align}\label{eq:mtexp}
  d\hat{\sigma}_{\text{exp},N} = \sum_{\rho=0}^N d\hat{\sigma}^{(\rho)} \left(\frac{\Lambda}{m_t}\right)^{2\rho},
\end{align}
where $\Lambda\in\left\{\sqrt{\hat{s}}, \sqrt{\hat{t}}, \sqrt{\hat{u}}, m_h\right\}$,
and determine the first few terms (up to $N=3$) of this asymptotic series
with the help of {\sc qgraf}~\cite{Nogueira:1991ex}, {\sc q2e}/{\sc
  exp}~\cite{Harlander:1997zb,Seidensticker:1999bb} 
and {\sc Matad}~\cite{Steinhauser:2000ry}, as well as {\sc Reduze}~\cite{vonManteuffel:2012np} and {\sc Form}~\cite{Vermaseren:2000nd,Kuipers:2012rf}.

We  applied the series expansion to the virtual corrections, 
combined with the infrared insertion operator ${\bom I}$, such that the expression in brackets below is infrared finite, 
  \begin{align}\label{eq:VI}
    &d\hat{\sigma}^{\mathrm{virt}} + d\hat{\sigma}^{\mathrm{LO}}(\eps) \otimes {\bom I}\nn\\
    &\approx 
    \left(d\hat{\sigma}_{\text{exp},N}^{\mathrm{virt}} + d\hat{\sigma}^{\mathrm{LO}}_{\text{exp},N}(\eps) \otimes {\bom I}\right) 
    \frac{d\hat{\sigma}^{\mathrm{LO}}(\eps)}{d\hat{\sigma}_{\text{exp},N}^{\mathrm{LO}}(\eps)} \;,
  \end{align}
such that we can set $\eps=0$ in
$d\hat{\sigma}^{\mathrm{LO}}/d\hat{\sigma}_{\text{exp},N}^{\mathrm{LO}}$.
  There is some freedom when to do the rescaling, \ie before/after the phase-space integration and 
  convolution with the PDFs. We opt to do it on a fully differential level, 
  \ie the rescaling is done for each phase-space point individually.
The comparison of this expansion with the full result is shown in the
next section.


\section{Numerical results}

In our numerical computation we set $\mur = \mu_F = \mu = m_{hh}/2$, 
 where $m_{hh}$ is the invariant mass of the Higgs boson pair.
We use the PDF4LHC15\_nlo\_100\_pdfas~\cite{Butterworth:2015oua,CT14,MMHT14,NNPDF} parton distribution functions,
along with the corresponding value for $\alpha_s$ for both the LO and
the NLO results. The masses have been set to $m_h=125$\,GeV, $m_t=173$\,GeV,
and the top-quark width has been set to zero.
We use a centre-of-mass energy of $\sqrt{s}=13$\,TeV 
and no cuts except a technical cut in the real radiation of
$p_{T}^{\rm{min}}=10^{-4}\cdot\sqrt{\hat{s}}$,
which we varied in the range $10^{-2} \leq
p_{T}^{\rm{min}}/\sqrt{\hat{s}} \leq 10^{-6}$ to verify that the
contribution to the total cross section is stable and independent of the cut within the numerical accuracy.

Including the top-mass dependence, we obtain the total cross section
at $\sqrt{s}=13$\,TeV
\begin{equation*}
\sigma^{\rm{NLO}}=27.80^{+13.8\%}_{-12.8\%}\, {\mathrm{fb}} \pm 0.3\%\, {\mathrm{(stat.)}} \pm 0.1\% \,\mathrm{(int.)}\;.
\end{equation*}
In addition to the dependence of the result on the variation of the scales by a factor of two around the central scale, we state the statistical error
coming from the limited number of phase-space points evaluated and the error stemming from the numerical
integration of the amplitude. The latter value has been obtained using error propagation and assuming Gaussian distributed errors and no correlation between the amplitude-level results.
The value of the cross section is 14\% smaller than the Born-improved
HEFT result, $\sigma^{NLO}_{HEFT}=32.22^{+18\%}_{-15\%}
\,\mathrm{fb}$, and about 40\% larger than the leading order result,
$\sigma^{LO} = 16.72^{+28\%}_{-21\%}\,\mathrm{fb}$.
Let us note that using a leading order PDF set rather
  than an NLO one for the LO calculation increases the LO result by
  about 10\%.

\begin{figure}[htb]
\includegraphics[width=0.5\textwidth]{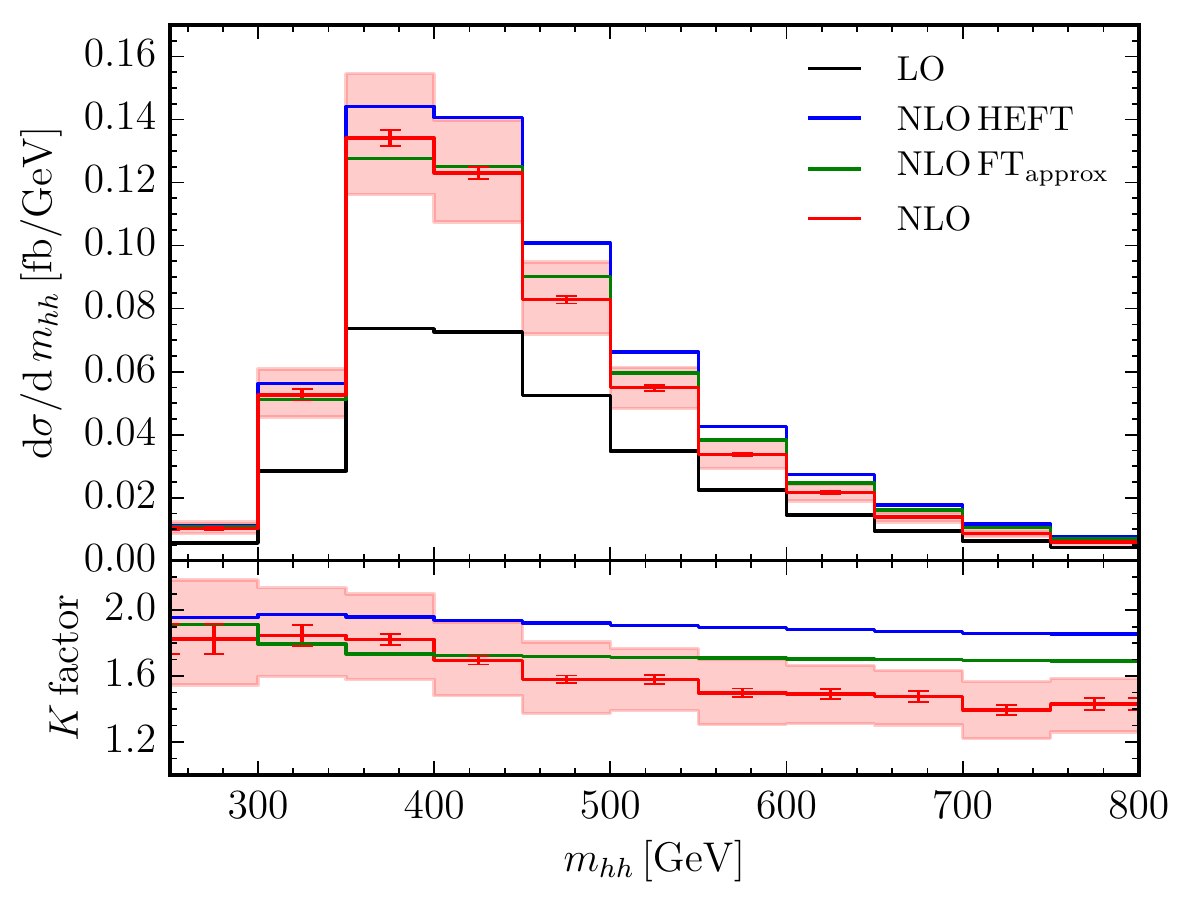}
\caption{Comparison of the full calculation to various
approximations for the Higgs pair invariant mass distribution at $\sqrt{s}=13$\,TeV.
``NLO HEFT'' denotes the effective field theory result, i.e
approximation (i) above, while ``FT$_{approx}$'' 
stands for approximation (ii), where the top-quark mass is taken into
account in the real radiation part only. The band results from scale
variations by a factor of two around the central scale $\mu=m_{hh}/2$.\label{fig:fullresult}}
\end{figure}

The results for the  $m_{hh}$ distribution are shown in Fig.~\ref{fig:fullresult}.
We can see that for $m_{hh}$ beyond $\sim 450$\,GeV, the top-quark mass effects lead to a
reduction of the $m_{hh}$ distribution 
by about 20-30\% as compared to the Born-improved HEFT approximation.
We also observe that the central value of the Born-improved HEFT result lies
outside the NLO scale uncertainty band of the full result for  $m_{hh}
\gtrsim 450$\,GeV, while the FT$_{approx}$ result, where the real
radiation contains the full mass dependence, lies outside the scale
uncertainty band  for $m_{hh}$ beyond $\sim 550$\,GeV. 
The scale uncertainty of the Born-improved HEFT and FT$_{approx}$ does not enclose the central value of the full
result in the tail of the $m_{hh}$ distribution.

\begin{figure}[htb]
\includegraphics[width=0.5\textwidth]{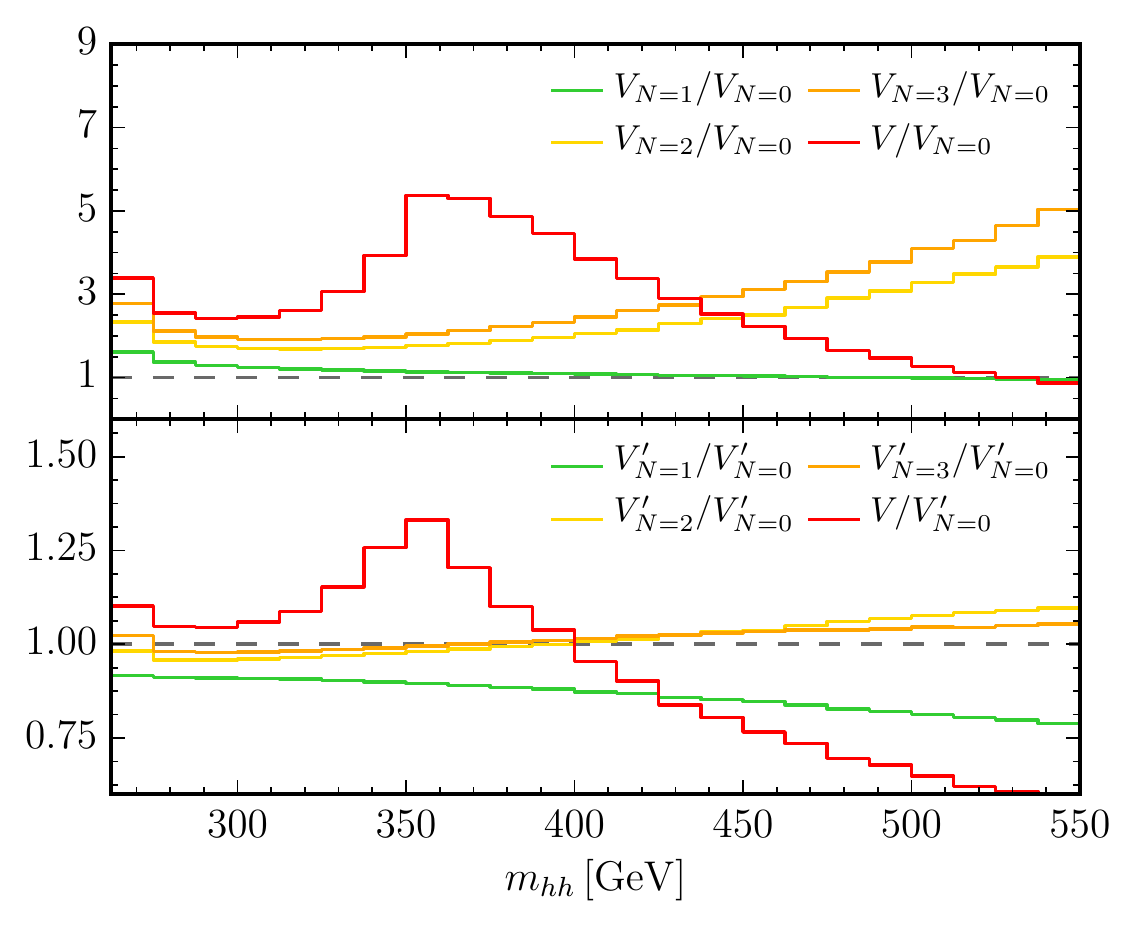}
\caption{Comparison of the virtual amplitude with full top-quark mass dependence to
  various orders in a $1/m_t^{2}$ expansion. $V^\prime_N$ denotes the
  Born-improved HEFT result to order $N$ in the $1/m_t^{2}$ expansion, \ie $V^\prime_N=V_N\,B/B_N$. \label{fig:virt}}
\end{figure}

In Fig.~\ref{fig:virt}, we show the results for the renormalized virtual amplitude 
including the $\mathbf{I}$-operator as defined in Ref.~\cite{Catani:1996vz} and
compare it to various orders in an 
expansion in $1/m_t^{2}$, see Eqs.~(\ref{eq:mtexp}),(\ref{eq:VI}).
In the upper panel  we normalize to the virtual HEFT result, while in
the lower panel we
normalize to the Born-improved HEFT result,
\ie $V^\prime_N=V_N\,B/B_N$.
The upper panel  shows that the agreement of the full result with the HEFT
result is only good well below the threshold at $2\,m_t$. The lower one
demonstrates that the deviations between the full result and the
Born-improved HEFT result are more than 30\% for $m_{hh}
\gtrsim 480$\,GeV.

\section{Conclusions}

We have calculated  the total cross section and the $m_{hh}$  distribution for Higgs boson pair production in gluon fusion at NLO, 
including the full top-quark mass dependence.
We have also presented results for the Born-improved HEFT (Higgs Effective Field
Theory) approximation, for the approximation where the virtual part is calculated in the 
Born-improved HEFT approximation while the real radiation part contains the 
full top-quark mass dependence (FT$_{approx}$), and for an expansion in $1/m_t^2$.
We observe that
the total cross section including the full top-quark mass dependence
is about 14\%  smaller than the one obtained within the Born-improved
HEFT approximation. 
The $m_{hh}$ distribution shows that for $m_{hh}$ values beyond $\sim 500$\,GeV,  the top quark mass effects lead
to a reduction of the differential cross section 
by about 20-30\%  as compared to the Born-improved HEFT approximation,
and by about 10-20\% as compared to the FT$_{approx}$ result.
Our results demonstrate that the calculation of the full top-quark
mass dependence is vital in order to get reliable predictions for
Higgs boson pair production over the full invariant mass range.

The method outlined here can in principle also be applied to the
calculation of other multi-scale amplitudes beyond one loop.

\vspace*{6mm}

\begin{acknowledgements}
\noindent{\it Acknowledgements}\\
We  would like to thank Stefano Di Vita, Thomas Hahn, Stephan Jahn, Gionata Luisoni, 
Pierpaolo Mastrolia, Anton Stoyanov and Valery Yundin for useful discussions.
We are also grateful to Andreas von Manteuffel for helpful
suggestions about the use of {\sc Reduze}.
This research was supported in part by the 
Research Executive Agency (REA) of the European Union under the Grant Agreement
PITN-GA2012316704 (HiggsTools) and the National Science Foundation under Grant No. NSF PHY11-25915.
SB acknowledges financial support by the ERC
Advanced Grant MC@NNLO (340983).
NG was supported by the Swiss National Science Foundation under contract
PZ00P2\_154829.
GH would like to acknowledge the Mainz Institute for Theoretical
Physics (MITP) for its hospitality. 
We gratefully acknowledge support and resources provided by the Max Planck Computing and Data Facility (MPCDF).
\end{acknowledgements}

 

\input{HH_PRL_corr.bbl}

\end{document}

%% file: definitions.tex
\newcommand{\als}{\ensuremath{\alpha_s}}

\newcommand{\mur}{\ensuremath{\mu_R}}

\newcommand{\msbar}{\ensuremath{\overline{\mathrm{MS}}}}

\def\eps{\epsilon}
\def\shat{\hat{s}}

\def\be{\begin{equation}}
\def\ee{\end{equation}}
\def\bea{\begin{eqnarray}}
\def\eea{\end{eqnarray}}
\def\nn{\nonumber}

\def\bom#1{{\mbox{\boldmath $\mathrm{#1}$}}}
\def\gsim{\raise0.3ex\hbox{$>$\kern-0.75em\raise-1.1ex\hbox{$\sim$}}}

\newcommand*{\eg}{e.g.\@\xspace}
\newcommand*{\ie}{i.e.\@\xspace}

%% file: HH_PRL_corr.bbl
%